\documentclass[prl,twocolumn,showpacs,superscriptaddress,floatfix]{revtex4}
\usepackage{graphicx}
\usepackage{float}
\begin{document}
\title{Theory of extraordinary transmission of light through quasiperiodic arrays of subwavelength holes}
\author{J. Bravo-Abad}
\affiliation{\mbox{Departamento de F\'{\i}sica Te\'{o}rica de la Materia Condensada
, Universidad Aut\'onoma de Madrid, E-28049 Madrid, Spain}}
\author{A.I. Fern\'andez-Dom\'{\i}nguez}
\affiliation{\mbox{Departamento de F\'{\i}sica Te\'{o}rica de la
Materia Condensada , Universidad Aut\'onoma de Madrid, E-28049
Madrid, Spain}}
\author{F.J. Garc\'{\i}a-Vidal}
\affiliation{\mbox{Departamento de F\'{\i}sica Te\'{o}rica de la
Materia Condensada , Universidad Aut\'onoma de Madrid, E-28049
Madrid, Spain}}
\author{L. Mart\'{i}n-Moreno}
\affiliation{\mbox{Departamento de F\'{\i}sica de la Materia Condensada, ICMA-CSIC,
 Universidad de Zaragoza, E-50009 Zaragoza, Spain}}
\begin{abstract}
By using a theoretical formalism able to work in both real and
k-spaces, the physical origin of the phenomenon of extraordinary
transmission of light through quasi-periodic arrays of holes is
revealed. Long-range order present in a quasiperiodic array
selects the wavevector(s) of the surface electromagnetic mode(s)
that allows an efficient transmission of light through
subwavelength holes.
\end{abstract}
\date{\today}
\maketitle

The phenomenon of extraordinary optical transmission (EOT) through
periodic two-dimensional (2D) arrays of subwavelength holes milled
in a metallic film \cite{Ebbesen98} has sparked a great deal of
interest due to both its fundamental implications and its broad
range of potential applications. Subsequent experimental and
theoretical works have concentrated on analyzing periodic
structures
\cite{Martin-Moreno01,Gomez-Rivas03,Gordon04,Barnes04,Klein04,Bravo-Abad04,Shen05,
Lalanne05,Hibbins06}. However, very recently, several experimental
studies showing EOT in quasiperiodic arrays of holes have been
reported \cite{Sun06,Schwanecke06,Przybilla06,Matsui07}. These
results suggest that the presence of long-range order in a 2D hole
array is the key ingredient to observe EOT.

In this Letter we present a complete physical explanation of the
EOT properties found in quasiperiodic distributions of
subwavelength holes. This analysis is based on the comparison
between the transmission properties of finite Penrose lattices of
holes with those associated with periodic arrays. The picture that
emerges from our theoretical study is that the physical origin of
EOT is common for both periodic and quasiperiodic arrays. It
relies on the excitation of surface electromagnetic (EM) modes
decorating the metallic interfaces.

\begin{figure}[h]
\begin{center}
\includegraphics[width=\columnwidth]{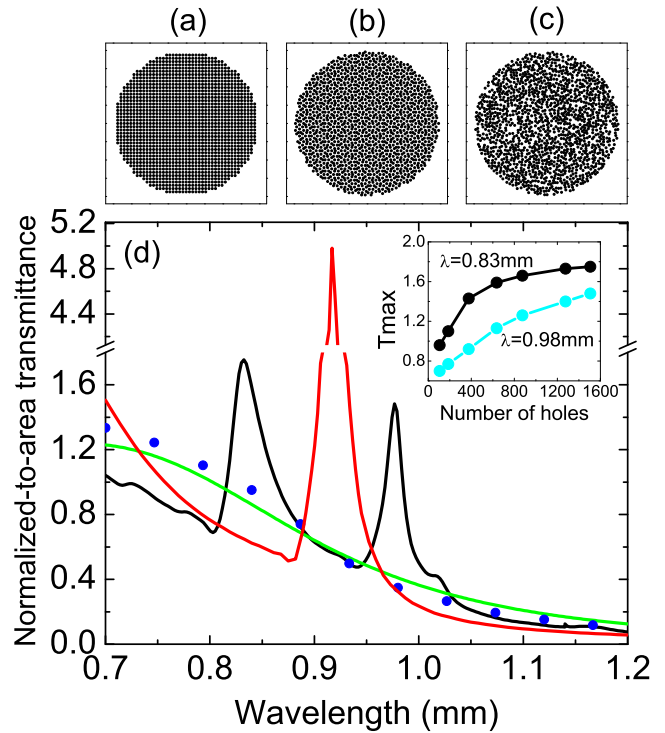}
\end{center}
\vspace*{-0.6cm} \caption{(color-on-line) (a-c) Structures
considered in this work. Square (left), Penrose (center) and
random lattices (right). (d) Normalized-to-area transmittance
($T$) spectra for: single hole (green line), square array (red
line), Penrose lattice (black line) and a random configuration
(blue dots). In all four cases, $a=0.2$mm, $h=0.075$mm and
$N=1506$. Inset in panel (d) shows the dependence with $N$ for $T$
at resonant peaks for the quasiperiodic array, $\lambda=0.83$mm
(black dots) and $\lambda=0.98$mm (cyan dots).}
\end{figure}

In this paper, our study is focused on analyzing the transmission
properties of Penrose lattices exhibiting ten-fold rotational
symmetry, as those studied experimentally in Ref. \cite{Matsui07}.
As in the experimental structure, the hole radius is chosen to be
$a=0.2$mm, the thickness of the metallic film is $h=0.075$mm and
the length of the rhombus side defining the Penrose tiling, $d$,
is $d=1$mm. In our simulations, metal is treated as a perfect
conductor (i.e. with dielectric constant $\epsilon= - \infty$),
which is an excellent approximation in the THz regime. In order to
calculate the scattering properties and EM field distributions, we
use a formalism based on a modal expansion of the fields at the
hole openings \cite{Bravo-Abad04}, which allows treating
efficiently large numbers of indentations, arbitrarily placed in a
metal film.

Figure 1 shows the three different types of hole arrangements
considered in this work. Left, center and right panels correspond
to a periodic square lattice, a ten-fold Penrose lattice and a
random distribution of circular holes, respectively. In all three
cases, the number of holes ($N=1506$), their diameter, the film
thickness and the external radius of the circular array are the
same. The coordinates in the Penrose lattice were generated by the
Dual Generalized Method \cite{Levine86,Rabson91}. The periodic
structure is a circular portion of a square lattice with lattice
parameter $P=0.89$mm. In the disordered case, holes are randomly
distributed but without allowing any interhole distance to be
smaller than the minimum one found in the quasiperiodic case.

Figure 1d depicts the normal incidence transmission spectra
computed for the three structures, along with the transmittance
associated with a single hole (green line). In all cases, the
transmittance is normalized to the flux of light impinging on the
area occupied by the holes. In the spectral range considered, the
single hole transmittance is a smooth decreasing function of the
wavelength. In the ordered case (red line), the transmittance
spectrum is also smooth, except close to the resonant peak
appearing at $\lambda=0.92$mm, where the normalized-to-area
transmittance ($T$) is about $5$ for the geometrical parameters we
are considering. This is the canonical EOT peak, appearing at a
resonant wavelength slightly larger that the lattice parameter. As
in the experiments, resonant transmission also appears when holes
are arranged in a Penrose lattice (black curve in Fig. 1). In this
case, maximum transmission values of about $1.5$ are obtained at
two resonant wavelengths, $\lambda=0.83$mm and $\lambda=0.98$mm.
The agreement between theory and experiment is remarkable (see
Fig. 2c in Ref.\cite{Matsui07}). On the other hand, blue dots in
Fig. 1d demonstrate that EOT does not appear for any distribution
of holes: the transmission spectrum for the random array does not
show any resonant feature. This is just a representative example
of disordered arrays; we have generated several random
configurations finding always a non-resonant behavior.

The appearance of EOT can be related to the lattice structure in
reciprocal space by extending arguments borrowed from the ordered
case to different lattices, as follows. Following Ref.
\cite{Bravo-Abad04}, EM fields in all space can be expressed in
terms of the modal amplitudes of the waveguide modes right at the
opening and the exit of the different holes ($E_\alpha({\bf R})$
and $E'_\alpha({\bf R})$, respectively, with ${\bf R}$ referring
to the 2D array locations and $\alpha$ running over the modes
inside the holes). These quantities can, in turn, be obtained by
solving a coupled system of equations. In order to find the link
between EOT and the structure factor of a given set of holes,
$S({\bf q})=\sum_{{\bf R}} \exp(-i{\bf q}{\bf R})$, it is
convenient to work with the Fourier components $E_\alpha({\bf q})
= \sum_{{\bf R}} \exp(- \imath{\bf q} {\bf R}) E_\alpha({\bf R})$,
which satisfy,
\begin{eqnarray}
(\Sigma_n({\bf q})-\epsilon_n) E_n({\bf q})- G_n^V E'_n({\bf q})&
=& I_{n} S({\bf q}-{\bf k}_0) \nonumber \\
 (\Sigma^{\prime}_n({\bf q})-\epsilon_n)E'_n({\bf q})- G_n^V
E_n({\bf q})& = & 0 \label{system}
\end{eqnarray}
where:
\begin{eqnarray}
\Sigma^{(\prime)}_n({\bf
q})=\frac{1}{E^{(\prime)}_n(\bf{q})}\sum_m \int d{\bf k} \:
G_{mn;{\bf k}} \: S({\bf q}-{\bf k}) \: E^{(\prime)}_m({\bf k})
\label{sigmaq}
\end{eqnarray}

The expression for the different quantities can be
straightforwardly obtained from the ones given in Ref.
\cite{Bravo-Abad04}. External illumination originates $I_{n}
S({\bf q}-{\bf k}_0)$, ${\bf k}_0$ being the in-plane component of
the incident wavevector. The term $\epsilon_n= -i Y_n
(1+\Phi_n)/(1-\Phi_n)$, with $\Phi_n \equiv \exp(2 \imath q_n h)$
($Y_n$ and $q_n$ being the admittance and the propagation constant
of mode $n$ inside the holes, respectively), is related to the
bouncing back and forth of EM fields inside the hole. $G_n^V= -2 i
Y_n \sqrt{\Phi_n}/(1-\Phi_n)$ couples the input and exit sides of
the hole. These magnitudes ($\epsilon_n$ and $G_n^V$) show no
dependence on parallel momentum, ${\bf k}$, as they do not couple
modes in different holes and they are real quantities for
subwavelength holes. The terms $\Sigma^{(\prime)}_n({\bf
q})E^{(\prime)}_n({\bf q})$ represent the scattering process that
couples $E^{(\prime)}_n({\bf q})$ to the continuum
$E^{(\prime)}_m({\bf k})$, the momentum difference being provided
by the lattice through $S({\bf q}-{\bf k})$. The amplitude of the
process depends on $G_{mn; {\bf k}}$:
\begin{equation}
G_{mn; {\bf k}}=\frac{i}{(2\pi)^2}\sum_{\sigma}Y_{{\bf
k}\sigma}<n|{\bf k}\sigma><{\bf k}\sigma|m>
\end{equation}
\noindent where the admittance of the plane wave ${\bf k}\sigma$,
$Y_{{\bf k}\sigma}$, is $g/k_z({\bf k})$ for a p-polarized wave
and $k_z({\bf k})/g$ for a s-polarized one, with $g=2\pi/\lambda$.
An important property that can be extracted from Eq. (3) is that
$G_{mn; {\bf k}}$ diverges whenever a p-polarized diffraction wave
goes glancing ($k_z=0$).
\begin{figure}[htb]
\begin{center}
\includegraphics[width=\columnwidth]{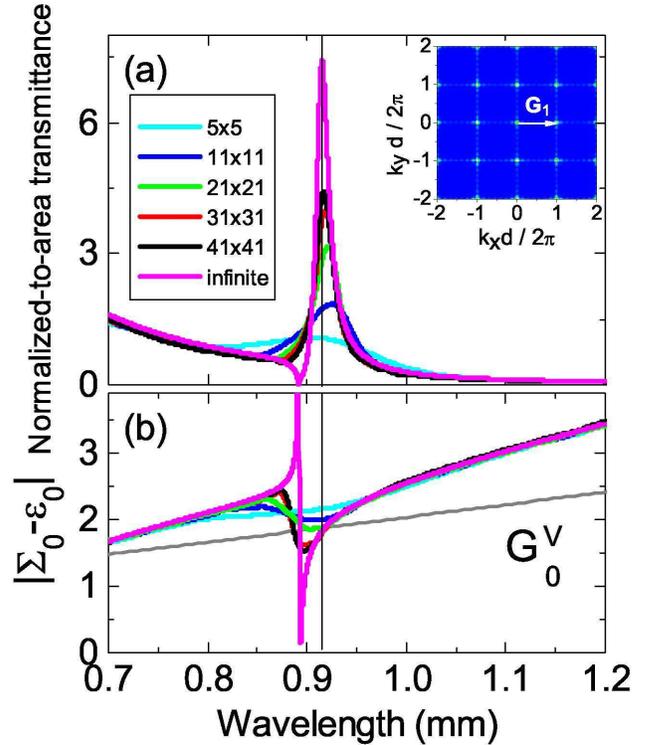}
\end{center}
\vspace*{-0.6cm} \caption{(color-on-line) (a) Normalized-to-area
transmittance versus wavelength for an infinite periodic array
(magenta line) and several finite square arrays. The geometrical
parameters are: $a=0.2$mm, $h=0.075$mm and $P=0.89$mm. Inset shows
the structure factor for the $41$x$41$ case. (b)
$|\Sigma_0-\epsilon_0|$ and $G^V_0$ versus wavelength for the
cases depicted in (a).}
\end{figure}

In order to illustrate the mathematics of the formation of surface
EM modes, let us consider the simpler system of an infinite
periodic array of holes. Additionally, we consider normal
incidence and assume that only one waveguide mode couples directly
to external radiation (the least evanescent mode, labelled as
$n$=0). By taking advantage of Bloch's theorem [$E_0({\bf k}+{\bf
G}_i)=E_0({\bf k})$ and $S({\bf k})=\sum_{i}\delta({\bf k}-{\bf
G}_i)$, being ${\bf G}_i$ a reciprocal lattice vector], Eqs. (1)
for ${\bf k}={\bf k}_0={\bf 0}$ transform into two simple
equations for $E_0({\bf 0})$ and $E'_0({\bf 0})$:
\begin{eqnarray}
(\Sigma_0-\epsilon_0) E_0({\bf 0})- G_0^V E'_0({\bf 0})& = &
I_{0} \nonumber \\
(\Sigma_0-\epsilon_0) E'_0({\bf 0})- G_0^V E_0({\bf 0})&= & 0
\end{eqnarray}
where $\Sigma_0=\sum_{{\bf G}_i}G_{00;{\bf G}_i}$. In Figure 2,
$T$ (panel a) and $|\Sigma_0-\epsilon_0|$ (panel b) versus
wavelength are depicted for an infinite periodic array (magenta
line). The geometrical parameters of this array are the same as
the periodic one analyzed in Fig.1. As $\Sigma_0$ for ${\bf G}=\pm
{\bf G}_1$ diverges at $\lambda=P=0.89$mm, both $E_0({\bf 0})$ and
$E'_0({\bf 0})$ are zero leading to zero transmission. This is the
so-called Wood's anomaly \cite{Ebbesen98} or anti-resonance as
quoted in Ref.\cite{Matsui07}. The crucial point to realize is
that, due to its rapid variation close to the divergence, at a
wavelength slightly larger than the one corresponding to glancing
angle, $|\Sigma_0-\epsilon_0|=G_0^V$. This leads to a resonant
enhancement of the electric field amplitudes at the interfaces of
the system [see Eqs.(4)], which can be assigned to the excitation
of a leaky surface EM mode \cite{Boardman}. Consequently, $T$
presents a maximum at the corresponding wavelength (see Fig.2).
Importantly, this resonance appears through the coupling to
p-polarized modes, closely resembling the EM fields of surface
plasmons in a real metal. Due to that, these modes are usually
called {\it spoof} surface plasmons emerging when the surface of a
perfect conductor is periodically corrugated \cite{Pendry04}.

\begin{figure}[htb]
\begin{center}
\includegraphics[width=\columnwidth]{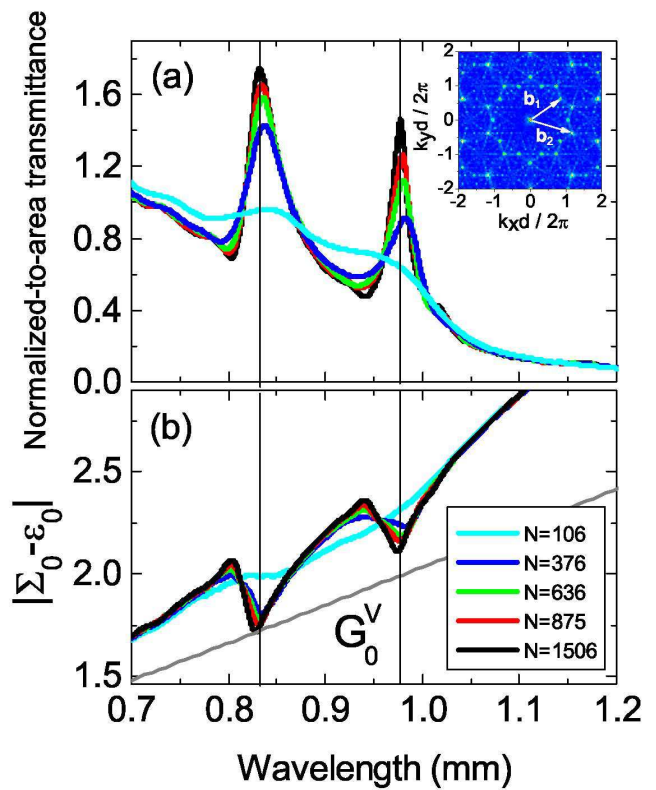}
\end{center}
\vspace*{-0.6cm} \caption{(color-on-line)(a) Normalized-to-area
transmittance versus wavelength for several quasiperiodic arrays
with different number of holes, $N$. The geometrical parameters
are: $a=0.2$mm, $h=0.075$mm and $d=1$mm. Inset shows the structure
factor for the $N=1506$ case. (b) Both $|\Sigma_0-\epsilon_0|$ and
$G_0^V$ versus wavelength for the cases depicted in (a).}
\end{figure}

The arguments presented above can be extended to the case of
finite arrays. Now Bloch's theorem cannot be applied and the
system of Eqs.(1) must be solved for a continuum of states ${\bf
q}$. However, we have checked that for a finite array with a large
number of holes ${\bf q}={\bf k}_0={\bf 0}$ is the dominant
transmission channel and the equations for $E_0({\bf 0})$ and
$E'_0({\bf 0})$ could be written like Eqs.(4) with $\Sigma_0$
numerically calculated from the knowledge of $E_0({\bf k})$ and
$E'_0({\bf k})$ (see Eq.(2)). The results of this approach are
shown in Figure 2 for the case of square periodic arrays of holes
(going from $5$x$5$ to $41$x$41$). In this case $\Sigma_0$
presents no divergences but resonant features appearing close to
$\lambda=P$. The first consequence is that Wood's anomalies do not
reach zero-value in finite arrays. As for the infinite case, the
cut between $|\Sigma_0-\epsilon_0|$ and $G_0^V$ marks the location
of the transmission peak for large arrays ($41$x$41$ and
$31$x$31$). For smaller arrays, there is no cut and the
transmission peak appears at the wavelength in which the {\it
distance} between $|\Sigma_0-\epsilon_0|$ and $G_0^V$ is minimal.
\begin{figure*}[htb]
\begin{center}
\includegraphics[width=17cm]{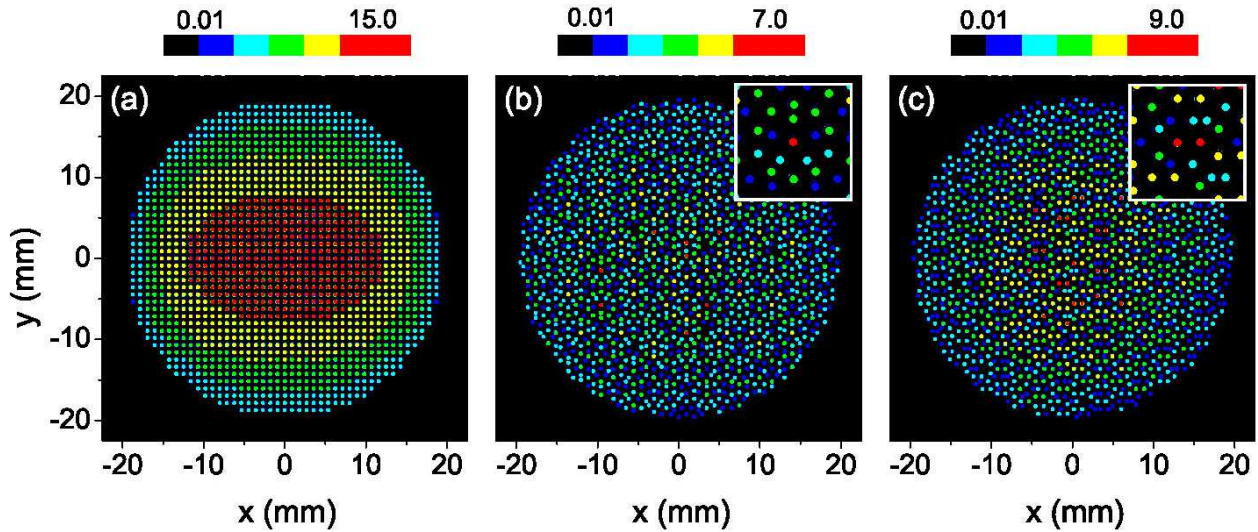}
\end{center}
\vspace*{-0.8cm} \caption{(color) Transmission per hole
(normalized to the single hole transmission) displayed in a color
scale for (a) ordered case evaluated at $\lambda=0.92$mm, (b)
Penrose lattice at $\lambda=0.83$mm and (c) Penrose lattice at
$\lambda=0.98$mm. The geometrical parameters are the same as in
Figure 1.}
\end{figure*}

Once described the periodic case, it is straightforward to analyze
the case of a quasiperiodic array of holes. In panel (b) of Figure
3, the evolution of $|\Sigma_0-\epsilon_0|$ versus wavelength is
studied for Penrose lattices with increasing number of holes
(ranging from $N=106$ to $N=1506$, the case analyzed in Fig.1).
$|\Sigma_0-\epsilon_0|$ present maxima at wavelengths
corresponding to the glancing condition for the two main
wavevectors of the structure factor (see inset of Fig.3a):
$\vec{b}_1$ ($\lambda_1=0.8$mm) and $\vec{b}_2$
($\lambda_2=0.94$mm). Consequently, $T$ shows two minima at these
two wavelengths. At slightly larger wavelengths, the distance
between $|\Sigma_0-\epsilon_0|$ and $G_0^V$ is minimal and,
correspondingly, two transmission peaks appear in the spectrum.
Therefore, these resonant transmission peaks stem from the
excitation of surface EM modes at the metallic surfaces, much in
the same way as in periodic arrays. Notice that, however, in the
quasiperiodic case, there is no minimum wavevector for diffraction
(i.e. the structure factor is non-zero for wavevectors with moduli
smaller than $|\vec{b}_1|$, see inset of Fig.3a). This results in
diffraction onto additional propagating modes in vacuum (other
than the zero-order mode), which leads to both smaller resonant
peaks and less pronounced Wood's anomalies than those emerging in
the periodic case.

It is worth analyzing how is the spatial distribution of light
emerging from the quasiperiodic array. Figure 4 renders the
transmission-per-hole in a Penrose lattice of $N=1506$ holes at
the two resonant wavelengths ($\lambda=0.83$mm and
$\lambda=0.98$mm in panels (b) and (c) of Fig.4, respectively).
For comparison, panel (a) of Fig. 4 shows the corresponding
distribution for the ordered array at the resonant wavelength
$0.92$mm. In all three cases, incident {\bf E}-field is pointing
along the $x$-direction. In the ordered case, due to finite size
effects, the maximum transmission is located at the center of the
structure \cite{Bravo-Abad06}. In quasiperiodic arrangements, the
transmission-per-hole distribution presents a completely different
pattern: it is far from being uniform, showing the appearance of
some holes with high transmission ({\it hot-spots}), which are
highlighted in the insets of panels (b) and (c). Interestingly, in
the Penrose lattice, for a given resonant wavelength, hot spots
show similar local environment. However, the existence of hot
spots does not imply that EOT in quasiperiodic systems is
dominated by very localized resonant configurations of holes.
Calculations (not shown here) on finite clusters of holes centered
at the hot spots show an increase of transmittance as a function
of number of neighbors included in the cluster. This point is
reinforced by the fact that the resonant peaks observed in the
transmission spectra of finite Penrose lattices do not saturate
for small $N$ values (see inset of Fig. 1d). Both these results
are consistent with the interpretation based on extended leaky
surface EM modes described above.

In conclusion, by developing a k-space theoretical formalism, we
have been able to demonstrate that the resonant features observed
in the transmission spectra of 2D Penrose lattices of holes can be
explained in terms of the formation of surface EM modes at the
interfaces of the metal film. Furthermore, we have linked the
formation of these modes to the structure factor of the hole
arrays, enabling the understanding of the appearance of
extraordinary optical transmission in more general conditions.

Financial support by the Spanish MECD under grant BES-2003-0374
and contract MAT2005-06608-C02 is gratefully acknowledged.

\end{document}